\documentclass[twocolumn,aps,prl,amsmath,amssymb,showpacs,preprintnumbers]{revtex4}
\usepackage{graphicx}
\begin{document}

\preprint{IGC--08/4--3}

\title{How quantum is the big bang?}

\author{Martin Bojowald}
\email{bojowald@gravity.psu.edu}
\affiliation{Institute for Gravitation and the Cosmos,
The Pennsylvania State University, 104 Davey Lab, University Park,
PA 16802, USA}

\pacs{98.80.Qc, 04.60.Pp, 98.80.Bp}

\newcommand{\lP}{\ell_{\mathrm P}}
\newcommand{\vp}{\varphi}
\newcommand{\vt}{\vartheta}

\newcommand{\md}{{\mathrm{d}}}
\newcommand{\tr}{\mathop{\mathrm{tr}}}
\newcommand{\sgn}{\mathop{\mathrm{sgn}}}

\newcommand*{\R}{{\mathbb R}}
\newcommand*{\N}{{\mathbb N}}
\newcommand*{\Z}{{\mathbb Z}}

\begin{abstract}
When quantum gravity is used to discuss the big bang singularity, the
most important, though rarely addressed, question is what role genuine
quantum degrees of freedom play. Here, complete effective equations
are derived for isotropic models with an interacting scalar to all
orders in the expansions involved. The resulting coupling terms show
that quantum fluctuations do not affect the bounce much. Quantum {\em
correlations}, however, do have an important role and could even
eliminate the bounce. How quantum gravity regularizes the big bang
depends crucially on properties of the quantum state.
\end{abstract}

\maketitle

Quantum cosmology is expected to be important for achieving a complete
understanding of the big bang which classically is singular and
preceded by a point of diverging energy density. This question is not
only important for our fundamental understanding of the universe but
may also have potential implications for future observations. Although
the subject is conceptually difficult, related to the fact that one is
dealing with a wave function of the whole universe, there has been
recent progress. Loop quantum cosmology \cite{LivRev} has provided a
resolution of the singularity in isotropic models based on the
extendability of the wave function across the classical singularity
\cite{Sing,BSCG}. However, this underlying behavior of the wave
function does not easily reveal a geometrical picture of what kind of
space-time region may replace the singularity. Often, singularity
resolution is expected to occur in the form of a bounce, a minimum
volume which is non-zero and where the collapse of a universe would be
stopped and turned around by quantum effects. This is in fact the only
possibility to avoid an isotropic singularity by a well-defined
space-time picture. But the question remains how strongly quantum the
big bang phase behaves, and what role genuine quantum variables such
as fluctuations play.

In loop quantum cosmology, the model of a free, massless scalar is
by now well-understood. It does, in fact, exhibit a simple bounce
described by the effective Friedmann equation \cite{RSLoopDual}
\begin{equation} \label{EffFried}
 \left(\frac{\dot{a}}{a}\right)^2 = \frac{4\pi
   G}{3}\frac{p_{\phi}^2}{a^6} \left(1- \frac{4\pi G}{3}\mu^2
   \frac{p_{\phi}^2}{a^6}\right)
\end{equation}
for the scale factor $a$ and with the momentum $p_{\phi}$ of the
scalar. The length parameter $\mu\not=0$, as seen below, arises due to
the loop quantization and determines a critical density $\rho_{\rm
crit}=3/8\pi G\mu^2$ such that the universe bounces ($\dot{a}=0$) when
$\rho_{\rm free}=\frac{1}{2}a^{-6}p_{\phi}^2=\rho_{\rm crit}$. That
this effective Friedmann equation reliably describes the evolution of
a wave packet (with $a=\langle\hat{a}\rangle$) in this model has been
numerically tested in \cite{APS} and proven rigorously in
\cite{BouncePert}. The key feature in this proof is that the specific
model is solvable for a certain factor ordering of the Hamiltonian,
which implies the absence of quantum back-reaction. The system is thus
free, and only finitely many variables are coupled in equations of
motion, rather than infinitely many ones as usually in quantum
systems. Quantum fluctuations, in fact, seem to play no role in
(\ref{EffFried}).  This weak influence of quantum variables on
expectation values is the reason for the very smooth nature of the
bounce, but based on this model alone it remains unclear how
representative this feature is in general.

Here, we substantially generalize the discussion by allowing for the
presence of a potential of the scalar. We will derive a new effective
Friedmann equation with additional quantum interactions and present,
to all orders in the potential as well as quantum degrees of freedom,
clear conditions for when the smooth bounce persists. We will see that
quantum variables do play important roles. However, for the bounce it
is not quantum fluctuations which are crucial but correlations between
different variables. Thus, the form of the bounce depends on the
squeezing of the state of the universe.

We use a canonical quantization, based on the conjugate variables
$(\phi,p_{\phi})$ for scalar matter and $(V,{\cal H})$ for space-time
where $V:= a^3/4\pi G\mu$ is proportional to the spatial volume and
${\cal H}=\mu\dot{a}/a$ to the Hubble parameter. Here, $\mu$ has no
classical effect but will be significant in the quantization
\footnote{Different canonical variables can make
$\mu$ $a$-dependent, which influences the dynamics in a way mimicking
lattice refinements of spatially discrete states
\cite{InhomLattice,SchwarzN}. Here, we consider only one case because
other choices do not lead to qualitative changes.}. It is useful to
describe evolution relationally, i.e.\ we do not solve for functions
in proper time, whose derivative is denoted by the dot, but for
relations between the variables such as $V(\phi)$. Due to general
covariance, the coordinate time dependence would have no intrinsic
meaning, and relational variables are closer to information contained
in the wave function. We will thus be treating $\phi$ as a measure for
time, such that its momentum $p_{\phi}$ determines the
Hamiltonian. Expressing the Friedmann equation as
\begin{equation} \label{pphi}
 -p_{\phi}=H:= \sqrt{12\pi GV^2{\cal H}^2-
   32\pi^2G^2\mu^2 V^2W(\phi)}
\end{equation}
and then quantizing the result implies a Schr\"odinger equation
\footnote{There are some subtleties in this step: A general issue is
how to choose the sign of $H$. For the quantum theory, this is related
to the physical inner product and the question which superpositions of
states one allows. As discussed in \cite{BounceCohStates}, this
affects only initial values but not equations of motion and can be
ignored here.  While this procedure is thus exact for a free, massless
scalar, the potential introduces technical difficulties. First, $\phi$
would not serve globally as internal time because it would not be
monotonic in coordinates. Thus, only patches of space-time can be
considered in this way, which then have to be combined. Secondly, at
the quantum level the first order equation (\ref{Schroed}) is not
equivalent to a second order equation
$\hat{p}_{\phi}^2\psi=\hat{H}^2\psi$ which would result if the
Friedmann equation or the Hamiltonian constraint is
quantized. However, the correction $[\hat{p}_{\phi},\hat{H}]$ can
simply be viewed as a factor ordering term and is proportional to
$\hbar V^2 W'(\phi)/H$ which is small for a flat potential. Moreover,
the Hamiltonian of the linear equation (\ref{Schroed}) can be
corrected for this term.}
\begin{equation} \label{Schroed}
 -\hat{p}_{\phi}\psi = i\hbar\frac{\partial\psi}{\partial\phi}=
 \hat{H}\psi\,.
\end{equation}

Loop quantum cosmology has no operator for ${\cal H}$, but rather
requires us to use the exponentials $\exp(i{\cal H})$
\cite{IsoCosmo,Bohr}. This is inherited from loop quantum gravity
\cite{Revs} where only holonomies are represented but
not connection components. To realize the solvability of the
free system, we use the variable $J:=V\exp(i{\cal H})$ in addition to
$V$, which upon quantization satisy a linear algebra
\[
 [\hat{V},\hat{J}_{\pm}]=\hbar\hat{J}_{\mp}\quad,\quad
 [\hat{J}_+,\hat{J}_-]=2\hbar(2\hat{V}+\hbar)
\]
with $\hat{J}_{\pm}:=\hat{J}\pm\hat{J}^{\dagger}$. Then, one can
choose a free Hamiltonian $\hat{H}_{\rm free}=-i\sqrt{3\pi
G}\hat{J}_-$ which has the correct low-curvature limit (${\cal H}\ll
1$) and is linear, implying solvability.

We will not require a complete quantum representation to capture
dynamical effects in effective equations. The main quantum effect
consists in terms which change the equations of motion of expectation
values by coupling them to quantum variables. This feature is
well-known from the Ehrenfest theorem, which can be extended to a
general procedure of deriving effective equations in canonical
quantizations \cite{EffAc}: We parameterize a state by its quantum
variables
\begin{equation}
 G^{V^aJ_-^bJ_+^c}:= \langle
   (\hat{V}-\langle\hat{V}\rangle)^a
   (\hat{J}_--\langle\hat{J}_-\rangle)^b
   (\hat{J}_+-\langle\hat{J}_+\rangle)^c\rangle_{\rm Weyl}
\end{equation}
defined as expectation values of Weyl ordered products of basic
operators with $a+b+c\geq 2$. This includes, for instance, the
fluctuation $G^{VV}=(\Delta V)^2$ or the covariance $G^{J_-J_+}=
(\Delta J)^2- (\Delta\bar{J})^2$.

Quantum variables enter correction terms due to quantum
back-reaction, which occurs in equations of motion
\begin{equation} \label{expvaleom}
 \frac{\md}{\md\phi}\langle\hat{O}\rangle=
 \frac{\langle[\hat{O},\hat{H}]\rangle}{i\hbar}
\end{equation}
for any expectation value $\langle\hat{O}\rangle$. In general, the
commutator $[\hat{O},\hat{H}]$ is non-linear in basic operators
$\hat{V}$ and $\hat{J}_{\pm}$, such that its expectation value depends
non-trivially on quantum variables. An exception are free systems,
where basic operators together with the Hamiltonian form a linear
algebra. In the case of $\hat{H}_{\rm free}$, equations of motion for
$\langle\hat{V}\rangle$ and $\langle\hat{J}_{\pm}\rangle$ depend only
on these expectation values. Thus, quantum variables do not at all
influence the motion of expectation values: there is no quantum
back-reaction.

This changes if the potential is non-zero and (\ref{pphi}) is
non-linear in $V$ and $J_-$. The expansion $H=\sum_{k=0}^{\infty}H_k$
with
\[
 H_k:=-i\sqrt{3\pi G}J_- \left(\begin{array}{c}
     \frac{1}{2}\\k\end{array}\right)
 \left(\frac{32\pi G\mu^2 V^2W}{3J_-^2}\right)^k
\]
is our starting point for quantization in Weyl ordering, which avoids
the square root but brings in infinitely many interaction
terms. Correspondingly, (\ref{pphi}) is replaced by
$-\langle\hat{p}_{\phi}\rangle= \langle\hat{H}\rangle=:H_Q$ with the
quantum Hamiltonian
\begin{equation} \label{HQ}
 H_Q= \sum_{k=0}^{\infty}\left(H_k+ \sum_{n=2}^{\infty} \sum_{a=0}^n
    \frac{\partial^n H_k}{\partial V^a\partial J_-^{n-a}}
   \frac{G^{V^aJ_-^{n-a}}}{a!(n-a)!}\right)\,.
\end{equation}
This can be derived by Taylor expanding $H(V+\Delta
\hat{V},J_-+\Delta\hat{J}_-)$ in the $\Delta\hat{O}:= \hat{O}-O$
where, to simplify notation, we dropped explicit brackets on
expectation values. There are thus 
terms containing quantum variables coupled to expectation values. 
(Their infinite number reflects the non-locality in time of quantum
physics. Defined as the expectation value $\langle\hat{H}\rangle$,
$H_Q$ must be finite; the $n$-sum in (\ref{HQ}) is thus convergent or at
least asymptotic.)

In a semiclassical expansion, one would keep only a finite number of
orders $n$, starting with fluctuations and covariances at $n=2$. In
this way, effective equations can be derived systematically based on
semiclassicality properties of an evolving state. General conclusions
drawn in this letter, however, will be valid to all orders in quantum
variables and are thus insensitive to the requirement of having a
semiclassical state of a certain form. This is especially important
near the big bang, where we have to analyze the role of quantum
variables without assuming too much.

The equation of motion for $\langle\hat{V}\rangle$ is derived from the
general formula (\ref{expvaleom}) and then expanded in quantum
variables.  Using the procedure in \cite{EffAc}, this gives,
\begin{eqnarray} \label{dVdphi}
 \frac{\md V}{\md\phi} &=& -iJ_+ \left(\frac{\partial
   H}{\partial J_-}+ \sum_{n=2}^{\infty}
 \sum_{a=0}^n  \frac{\partial^{n+1}H}{\partial V^a\partial
   J_-^{n-a+1}} \frac{G^{V^aJ_-^{n-a}}}{a!(n-a)!}\right)\nonumber\\
 && -i \sum_{n=2}^{\infty} \sum_{a=0}^n
  \frac{\partial^n H}{\partial V^a\partial J_-^{n-a}}
 \frac{G^{V^aJ_-^{n-a-1}J_+}}{a!(n-a-1)!}
\end{eqnarray}
as the equation of motion. We will perform a perturbative analysis to
all orders in both expansions by the potential in $H$ and in quantum
variables. The latter corresponds to a construction valid to all
orders in a loop expansion via all $n$-point functions.
Moreover, we are perturbatively expanding around the loop-quantized
$\hat{H}_{\rm free}$, such that non-perturbative discreteness
contained in this Hamiltonian is realized. Thus, we are providing
properties of solutions to the fundamental difference equation.

By itself, (\ref{dVdphi}) cannot be solved because $V$ is coupled not
only to $J_{\pm}$ but also to all infinitely many quantum
variables. But we will see that statements about the bounce can
nevertheless be made at this general level. This will constitute our
main result.

To see this, we derive an effective Friedmann equation analogous to
(\ref{EffFried}) but valid also in the presence of a potential. This
requires us to eliminate the $J_{\pm}$-dependence from (\ref{dVdphi})
by using the quantum Hamiltonian (\ref{HQ}). To any finite order $k$
in the potential, the quantum Hamiltonian takes the form of
$J_-^{1-2k}$ multiplied with a polynomial of order $2k$ in
$J_-$. Thus, the equation $H_Q=-p_{\phi}$ can be solved for $J_-$ in
terms of only $V$, $W(\phi)$ and $p_{\phi}$. Solutions to the
polynomial equation can easily be found perturbatively, starting from
the root $iJ_-=p_{\phi}(1+a^6
W(\phi)(1+\epsilon_1)/p_{\phi}^2)/\sqrt{3\pi G}$ for $k=1$, where
\[
 \epsilon_1 = \sum_{n=2}^{\infty} (-1)^n
 \left(\frac{G^{J_-^n}}{J_-^n}- 2\frac{G^{VJ_-^{n-1}}}{VJ_-^{n-1}}+
 \frac{G^{V^2J_-^{n-2}}}{V^2J_-^{n-2}}\right)
\]
denotes a correction in terms of relative quantum variables.  In
(\ref{dVdphi}), also $J_+$ appears which we determine from $J_-$ using
$\frac{1}{4}\left(J_+^2+(iJ_-)^2\right)= J\bar{J}$.  With the identity
$\hat{J}\hat{J}^{\dagger}=\hat{V}^2$, we write $J\bar{J}$ as $
\frac{1}{2}\langle\hat{J}\hat{J}^{\dagger}+
\hat{J}^{\dagger}\hat{J}\rangle - G^{J\bar{J}} = (V+\hbar/2)^2+(\Delta
V)^2-G^{J\bar{J}}+\frac{\hbar^2}{4} $ and thus have
\[
 \frac{J_+}{2V+\hbar}= \pm\sqrt{1-\left(\frac{J_-}{i(2V+\hbar)}\right)^2-
   \epsilon_0}
\]
with $\epsilon_0=(G^{J\bar{J}}-(\Delta
V)^2-\hbar^2/4)/(V+\hbar/2)^2$.

Using the solution for $iJ_-$ to all orders, where analogously to
$\epsilon_1$ we have $\epsilon=\sum_k \epsilon_{k+1}
(a^6W/p_{\phi}^2)^k$ from (\ref{HQ}), the leading term of $J_+$ is
$J_+= \pm 2V\sqrt{1-\rho_Q/\rho_{\rm crit}}$ with
\begin{equation}\label{rhoQ}
 \rho_Q:=\rho+\epsilon_0 \rho_{\rm crit}+ W \sum_{k=0}^{\infty}\epsilon_{k+1}
 \left(\frac{a^6W}{p_{\phi}^2}\right)^k\,.
\end{equation}
Because $J_+$ is always real, this proves the {\em upper bound
$\rho_Q\leq\rho_{\rm crit}$ for arbitrary potentials.}  In
(\ref{dVdphi}), this gives a similar to what appears in the free
effective Friedmann equation (\ref{EffFried}). There are only
corrections to $\rho$ by relative quantum variables in
$\epsilon_k$. The first line of (\ref{dVdphi}) has only terms
containing $J_+$ as a pre-factor. Thus, even in the presence of a
potential the effective Friedmann equation does have a factor close to
$1-\rho/\rho_{\rm crit}$ which reduces $\dot{a}$ once energy density
reaches $\rho_{\rm crit}$. However, there are additional terms in
(\ref{dVdphi}) which do not depend on $J_+$ and thus do not obtain
such a factor. These terms must be discussed further before a general
conclusion about the bounce, where $\dot{a}=0$ exactly, can be
reached. With them, collectively denoted as $\eta$, the effective
Friedmann equation, derived with $3\dot{a}/a=\dot{\phi}V^{-1}\md
V/\md\phi$ and $\dot{\phi}=a^{-3}p_{\phi}$, takes the form (see
\cite{QuantumBounce} for some details)
\begin{eqnarray}
 \left(\frac{\dot{a}}{a}\right)^2 &=& \frac{8\pi G}{3}\left(\rho
 \left(1-\frac{\rho_Q}{\rho_{\rm crit}}\right)\right.\\
 &&\pm\left.\frac{1}{2}\sqrt{1-\frac{\rho_Q}{\rho_{\rm crit}}} 
\eta W+ \frac{a^6W^2}{2p_{\phi}^2}\eta^2
\right)\,. \nonumber
\end{eqnarray}
In $\eta=\sum_k\eta_{k+1}(a^6W/p_{\phi}^2)^k$ as in $\rho_Q$, we have
contributions from all powers of the potential. From the linear term,
for instance, we have from (\ref{dVdphi}):
\begin{eqnarray*}
 \eta_1 &=& \sum_{n=2}^{\infty} (-1)^n
 \left(n\frac{G^{J_-^{n-1}J_+}}{J_-^n}-
 2(n-1)\frac{G^{VJ_-^{n-2}J_+}}{VJ_-^{n-1}}\right.\\
&&\left.+
 (n-2)\frac{G^{V^2J_-^{n-3}J_+}}{V^2J_-^{n-2}}\right)\,.
\end{eqnarray*}

The new terms in (\ref{dVdphi}) and $\eta$ contain only
covariances since there is always a single
factor of $\hat{J}_+$ which must be accompanied by at least one other
basic operator for a non-zero quantum variable. Thus, {\em a state
which is completely uncorrelated when its corrected energy density
$\rho_Q$ reaches $\rho_{\rm crit}$ bounces at this time.} The free
value of $\rho_{\rm crit}$ is unchanged. This result holds {\em to all
orders in the potential and in the moment expansion} and thus
substantially generalizes the free result. However, the high-energy
behavior changes for $\eta\not=0$ in a way which affects the condition
$\dot{a}=0$.  This is difficult to evaluate because correlations are
dynamical, subject to equations of motion which just as
(\ref{expvaleom}) follow from commutators with the Hamiltonian.  For
instance, $\md G^{J_-J_+}/\md\phi$ couples to all quantum variables in
the presence of a potential and we have, e.g., a term $-16iV^3 (\Delta
V)^2W(\phi) /J_-^3$ which cannot be zero due to uncertainty relations
\cite{BouncePot}. Moreover, the pre-factor $V^3/J_-^3$ is large for
large volume since $J_-$, for free solutions, is proportional to the
bounce volume. Thus, correlations can grow significantly during long
evolution.

Correlations at the would-be bounce are related to an initial state in
a complicated way, especially since one requires long evolution times
to get to the would-be bounce from a semiclassical state at large
volume where one may make assumptions on the state. Even if $\eta$
remains small, it has to be compared with the small
$1-\rho_Q/\rho_{\rm crit}$ in the vicinity of a possible bounce. Thus,
its precise value as well as time dependence matter. The bounce
changes for squeezed states with non-zero correlations.

In this context, it is interesting to note that parameters which
determine the correlations are difficult to control even for the free
system \cite{BeforeBB}. For the free system, one can find
explicit solutions for expectation values and quantum variables
\cite{BounceCohStates}, and correlations turn out to vanish at the
free bounce for states whose fluctuations are symmetric around the
bounce. If such a state approximately decribes the bounce of the
interacting system, also the interacting system bounces at the same
critical density. However, this is only a special class of states, and
it is difficult, if not impossible, to determine whether our universe
would be in an unsqueezed or squeezed state. That the symmetry of
quantum variables around the bounce is relevant also for the
interacting system can be seen by a general argument: time reversal
around the bounce is implemented by the mapping $J_+\mapsto -J_+$
which is a reflection ${\cal H}\mapsto \pi-{\cal H}$ around the bounce
point ${\cal H}=\pi/2$ where $J_+=0$. Thus, any of the correlations
$G^{V^aJ_-^bJ_+}$ appearing in the second line of (\ref{dVdphi}) would
change sign for states which are time reversal symmetric around the
bounce. Correlations then vanish at the bounce and $\md V/\md\phi$ is
proportional to $J_+\propto \sqrt{1-\rho_Q/\rho_{\rm crit}}$ which
implies the bounce. For non-symmetric states, $\eta\not=0$, the
questions raised and addressed in \cite{BeforeBB} about
properties of quantum variables before the bounce become relevant even
to address the bounce itself.

In summary, we have generalized the effective Friedmann equation
previously available only for free models to {\em arbitrary scalar
matter which may be massive and self-interacting} in a spatially flat
isotropic space-time. Here, also pressure $P$ enters since
$W=\frac{1}{2}(\rho-P)$ and $a^{-6}p_{\phi}^2=\rho+P$ appear
separately. We have found specific conditions under which a bounce at
a critical value of the quantum corrected density $\rho_Q$ results,
but also highlighted the detailed role which the precise quantum state
has to play. In particular, the {\em squeezing of states}, which has
often been overlooked in this context, {\em does have a crucial
bearing on the form of the bounce.} If correlations have developed
strongly by the time when the energy density approaches the critical
value, they may prevent a bounce if no solution for $\dot{a}=0$
exists. Or, there may be several solutions if the complete dynamical
behavior of the squeezing parameter $\eta$ is taken into account. In
this case, the universe could be stuck in an oscillatory quantum mode
at small volume rather than opening up to another large classical
universe before the big bang. This interesting possibility would
resemble scenarios which have been developed by combining effects of
loop quantum cosmology with the emergent universe
\cite{Emergent}.  Which precise
possibility is realized is currently only a matter of speculation
since the dynamical squeezing must be brought under much more
control. Nevertheless, at a fundamental level the models considered
here are non-singular independently of the matter Hamiltonian as shown
in \cite{Sing} (and \cite{NonMin} for non-minimal coupling).

Despite some remaining inconclusiveness, we have demonstrated that the
methods used here, which present the first perturbative tools for loop
quantum gravity, are sufficiently well-developed to provide valuable
insights. They have allowed us to draw conclusions, such as an upper
bound for $\rho_Q$ in (\ref{rhoQ}), valid {\em to all orders in a
moment expansion of states.} This goes well beyond the semiclassical
approximation. We thus generalized results of free models
significantly, and replaced numerical simulations by analytic
calculations with full access to the parameter space. With systematic
ways to derive effective equations, which are being extended to
include inhomogeneities, the behavior of loop quantum gravity can be
studied in much more detail than presently available.

\bigskip

\noindent {\bf Acknowledgements:}
This work was supported in part by NSF grant PHY0653127.


\begin{thebibliography}{10}


\bibitem{LivRev}
M. Bojowald, Living Rev.\ Relativity {\bf 8},  11  (2005).

\bibitem{Sing}
M. Bojowald, Phys.\ Rev.\ Lett. {\bf 86},  5227  (2001).

\bibitem{BSCG}
M. Bojowald,  AIP Conf.\ Proc. {\bf 910}, 294 (2007).

\bibitem{RSLoopDual}
P. Singh, Phys.\ Rev.\ D {\bf 73},  063508  (2006).

\bibitem{APS}
A. Ashtekar, T. Pawlowski, and P. Singh, Phys.\ Rev.\ Lett. {\bf 96},  141301
  (2006); Phys.\ Rev.\ D {\bf 73},  124038
  (2006).

\bibitem{BouncePert}
M. Bojowald, Phys.\ Rev.\ D {\bf 75},  081301(R)  (2007).

\bibitem{IsoCosmo}
M. Bojowald, Class.\ Quantum Grav. {\bf 19},  2717  (2002).

\bibitem{Bohr}
A. Ashtekar, M. Bojowald, and J. Lewandowski, Adv.\ Theor.\ Math.\ Phys. {\bf
  7},  233  (2003).

\bibitem{Revs}
C. Rovelli, {\em Quantum Gravity} (Cambridge University Press, Cambridge, UK,
  2004);
A. Ashtekar and J. Lewandowski, Class.\ Quantum Grav. {\bf 21},  R53  (2004);
T. Thiemann, {\em Introduction to Modern Canonical Quantum General Relativity}
(Cambridge University Press, Cambridge, UK,
  2007).

\bibitem{EffAc}
M. Bojowald and A. Skirzewski, Rev.\ Math.\ Phys. {\bf 18},  713  (2006);
Int.\ J.\ Geom.\ Meth.\ Mod.\ Phys. {\bf 4},  25
   (2007).

\bibitem{QuantumBounce}
M. Bojowald, Gen.\ Rel.\ Grav., to appear, [arXiv:0801.4001].

\bibitem{BouncePot}
M. Bojowald, H. Hern\'andez, and A. Skirzewski, Phys.\ Rev.\ D {\bf 76},
  063511  (2007).

\bibitem{BeforeBB}
M. Bojowald, Nature Physics {\bf 3},  523  (2007);
Proc.\ Roy.\ Soc.\ A, to appear, [arXiv:0710.4919].

\bibitem{BounceCohStates}
M. Bojowald, Phys.\ Rev.\ D {\bf 75},  123512  (2007).

\bibitem{Emergent}
G.~F.~R. Ellis and R. Maartens, Class.\ Quant.\ Grav. {\bf 21},  223  (2004);
D.~J. Mulryne, R. Tavakol, J.~E. Lidsey, and G.~F.~R. Ellis, Phys.\ Rev.\ D
  {\bf 71},  123512  (2005);
M. Bojowald, Nature {\bf 436},  920  (2005);
L. Parisi, M. Bruni, R. Maartens, and K. Vandersloot, Class.\ Quantum Grav.
  {\bf 24},  6243  (2007).

\bibitem{NonMin}
M. Bojowald and M. Kagan, Class.\ Quantum Grav. {\bf 23},  4983  (2006).

\bibitem{InhomLattice}
M. Bojowald, Gen.\ Rel.\ Grav. {\bf 38},  1771  (2006).

\bibitem{SchwarzN}
M. Bojowald, D. Cartin, and G. Khanna, Phys.\ Rev.\ D {\bf 76},  064018
  (2007).

\end{thebibliography}
\end{document}